\author{Martin Radloff}
\pdfoutput=1
	\documentclass[a4paper,12pt,oneside]{scrartcl}
	
\usepackage[sectionbib]{natbib}
\usepackage[]{hyperref}
\renewcommand{\theequation}{\thesection\arabic{equation}}
	\usepackage[english]{babel}   
	\usepackage[latin1]{inputenc}
	\usepackage[T1]{fontenc}
	\usepackage{ifthen}
	\usepackage{amsmath}
	\usepackage{amssymb}
	\usepackage{amsthm}
	\usepackage{latexsym, mathrsfs}
	\usepackage{bbm}

	\usepackage{geometry}
	\usepackage{graphicx}
	\usepackage{subfigure}
	\usepackage{xcolor}
\usepackage{pgf,tikz}
\usetikzlibrary{arrows}	
	\usepackage{multirow}
	\usepackage{multicol}
	\usepackage{enumerate, enumitem}	
	\usepackage{url}
	\usepackage{scrlayer-scrpage}
	\usepackage{suetterl}	
\newfont{\suet}{suet14}
\newfont{\schwell}{schwell}
\DeclareTextFontCommand{\textsuet}{\suet}

\DeclareTextFontCommand{\textschwell}{\schwell}	

	\usepackage{datetime}
	\usepackage{blindtext}
	\defpagestyle{Kopfi}{
		{} {} {\small Martin Radloff, Rainer Schwabe\hfill Exact Designs on the Ball
		}
	}{
		{} {} {\hfill\pagemark}
	}

\usepackage{dsfont}
\usepackage{mathtools}
\DeclareMathOperator{\sign}{sign}
\usepackage[title]{appendix}%

\newcommand{\BIGOP}[1]{\mathop{\mathchoice%
{\raise-0.22em\hbox{\huge $#1$}}%
{\raise-0.05em\hbox{\Large $#1$}}{\hbox{\large $#1$}}{#1}}}

\newcommand{\BIGboxplus}{\mathop{\mathchoice%
{\raise-0.35em\hbox{\huge $\boxplus$}}%
{\raise-0.15em\hbox{\Large $\boxplus$}}{\hbox{\large $\boxplus$}}{\boxplus}}}

\newtheorem{theorem}{Theorem}
\newtheorem{lemma}{Lemma}

\theoremstyle{definition}
\newtheorem{definition}{Definition}

\newtheorem{remark}{Remark}

\newcommand{\D}{\mathrm{d}}
\newcommand{\id}{\operatorname{id}}

\usepackage{dsfont}

\begin{document}
	\pagestyle{Kopfi}
	\thispagestyle{empty}
	\vspace*{1.3cm}
	\begin{center}
		{\Large\textbf{$D$-Optimal and Nearly $D$-Optimal Exact Designs for Binary Response on the Ball}
		}
	\end{center}
	\centerline{Martin Radloff\footnote[2]{corresponding author: Martin Radloff, Institute for Mathematical Stochastics, Otto-von-Guericke-University, PF~4120, 39016~Magdeburg, Germany, \url{martin.radloff@ovgu.de}} and Rainer Schwabe\footnote[3]{Rainer Schwabe, Institute for Mathematical Stochastics, Otto-von-Guericke-University, PF~4120, 39016~Magdeburg, Germany, \url{rainer.schwabe@ovgu.de}}} 
		
	\vspace*{1.3cm}

%

\begin{quotation}
\noindent {\textit{Abstract:}}
In this paper the results of \cite{Radloff:2019:moda} will be extended for a special class of symmetrical intensity functions. This includes binary response models with logit and probit link. To evaluate the position and the weights of the two non-degenerated orbits on the $k$-dimensional ball usually a system of three equations has to be solved. The symmetry allows to reduce this system to a single equation. 
As a further result, the number of support points can be reduced to the minimal number. 
These minimally supported designs are highly efficient. 
The results can be generalized to arbitrary ellipsoidal design regions.

\vspace{9pt}
\par
\end{quotation}\par

\def\thefigure{\arabic{figure}}
\def\thetable{\arabic{table}}

\renewcommand{\theequation}{\thesection.\arabic{equation}}

\fontsize{12}{14pt plus.8pt minus .6pt}\selectfont

\setcounter{equation}{0} 

\section{Introduction}
\label{sec:intro}
	Spherical design spaces can occur in engineering or physics problems where the validity of a model may be assumed on a spherical region around a target value. So (linear) models on spherical design spaces were investigated early in publications like \cite{Kiefer:1961} and \cite{Farrell:1967} which discuss polynomial regression on the ball. These ideas were followed up by papers in which also only linear problems were focused. So \cite{Lau:1988} fitted polynomials on the $k$-dimensional unit ball by using canonical moments. In \cite{Dette:2005, Dette:2007} and \cite{Hirao:2015} harmonic polynomials and Zernike polynomials were used to be fit on the unit disc (2-dimensional unit ball), the 3- and $k$-dimensional unit ball. 	
	On the other hand generalized linear models are also well-examined and used in practical application. Logit and probit models, for example, in one dimension on an interval have already been investigated by \cite{Ford:1992} and \cite{Biedermann:2006}. But there seems to be no available literature which combines both topics. 
		
	In our publication \cite{Radloff:2019:jspi} we took the first step to bring non-linearity or generalized linear models, respectively, and spherical design regions together. These results were extended to a wider class of non-linear models in our follow-up paper \cite{Radloff:2019:moda}.
	
	For better comprehensibility, we will start with the model description and give a brief overview of the findings so far. Then we will consider a special class of intensity functions which allows to reduce the the complexity of finding (locally) $D$-optimal designs. Afterwards we will tackle the problem, that the optimal designs are not exact designs in general, by establishing highly efficient designs on the ball.

\section{General Model Description}
\label{sec:model}		
	As in \cite{Radloff:2019:jspi} and \cite{Radloff:2019:moda}, where we described (locally) $D$-optimal designs for two special classes of linear and non-linear models on a $k$-dimensional unit ball $\mathbb{B}_k=\{\boldsymbol{x}\in\mathbb{R}^k\ :\ x_1^2+\ldots+x_k^2\leq 1\}$ with~$k\in\mathbb{N}$, we solely focus (non-linear) multiple regression models, which means the linear predictor is
	\begin{equation*}
		\boldsymbol{f}(\boldsymbol{x})^\top\boldsymbol{\beta} = \beta_0 + \beta_1 x_1 + \ldots + \beta_k x_k\ 
	\end{equation*}
	with regression function $\boldsymbol{f}:\mathbb{B}_k\to\mathbb{R}^{k+1}$, $\boldsymbol{x}\mapsto(1,x_1,\ldots,x_k)^\top$, and parameter vector $\boldsymbol{\beta}=(\beta_0,\beta_1,\ldots,\beta_k)^\top\in\mathbb{R}^{k+1}$.
	The one-support-point (or elemental) information matrix should be representable in the form
	\begin{equation*}
		\boldsymbol{M}(\boldsymbol{x},\boldsymbol{\beta})=\lambda\!\left(\boldsymbol{f}(\boldsymbol{x})^\top\boldsymbol{\beta}\right)\boldsymbol{f}(\boldsymbol{x})\boldsymbol{f}(\boldsymbol{x})^\top
	\end{equation*}
	with an intensity (or efficiency) function $\lambda$ which only depends on the value of the linear predictor $\boldsymbol{f}(\boldsymbol{x})^\top\boldsymbol{\beta}$.	These one-support-point (or elemental) information matrices are the base for the information matrix of a (generalized) design $\xi$ with independent observations	
	\begin{equation*}
		\boldsymbol{M}(\xi,\boldsymbol{\beta})=\int\boldsymbol{M}(\boldsymbol{x},\boldsymbol{\beta})\ \xi(\D \boldsymbol{x})=\int\lambda\!\left(\boldsymbol{f}(\boldsymbol{x})^\top\boldsymbol{\beta}\right)\boldsymbol{f}(\boldsymbol{x})\boldsymbol{f}(\boldsymbol{x})^\top \xi(\D \boldsymbol{x})\ .
	\end{equation*}	
	Here generalized design means an arbitrary probability measure on the design region~$\mathbb{B}_k$.
	
	These information matrices allow to define the (local) $D$-optimality, which is one of the most popular criteria in experimental design theory. A design~$\xi_{\boldsymbol{\beta}^0}^\ast$ with regular information matrix $\boldsymbol{M}(\xi_{\boldsymbol{\beta}^0}^\ast,\boldsymbol{\beta}^0)$ is called (locally) $D$-optimal (at~$\boldsymbol{\beta}^0$) if $\det(\boldsymbol{M}(\xi_{\boldsymbol{\beta}^0}^\ast,\boldsymbol{\beta}^0))\geq\det(\boldsymbol{M}(\xi,\boldsymbol{\beta}^0))$ holds for all suitable probability measures $\xi$ on the design space --- here $\mathbb{B}_k$. This optimality criterion can be interpreted as the minimization of the volume of the (asymptotic) confidence ellipsoid.

\section{Prior Results}
\label{sec:prior}

	In \cite{Radloff:2016} we stated results on equivariance and invariance. By rotating the design space~$\mathbb{B}_k$ --- the $k$-dimensional unit ball --- and the parameter space $\mathbb{R}^{k+1}$ in an analogous way the linear predictor of the multiple regression problem reduces to 
		\begin{equation}
			\boldsymbol{f}(\boldsymbol{x})^\top\boldsymbol{\beta} = \beta_0 + \beta_1 x_1 \text{\quad and \quad $\beta_1\geq 0$}\ .
		\label{eq:simplifiedModel}
		\end{equation}
	Using the rotation invariance with fixed~$x_1$, this means the invariance to all orthogonal transformations in~$O(k)$ which let the $x_1$-component unchanged, the (locally) $D$-optimal (generalized) design~$\xi^\ast$ can be decomposed (\mbox{$\xi^\ast=\xi_1^\ast\otimes\overline{\eta}$}) in a marginal probability measure $\xi_1^\ast$ on $[-1,1]$ for $x_1$ and a probability kernel~$\overline{\eta}$ given $x_1$.	For fixed $x_1$ the kernel $\overline{\eta}(x_1,\cdot)$ is the uniform distribution on the surface of a $(k-1)$-dimensional ball with radius $\sqrt{1-x_1^2}$ --- the orbit at position~$x_1$.\\		
	As a consequence the multidimensional problem collapses to a one-dimensional marginal problem. Only the positions of the orbits and their weights have to be determined. To get an exact design the uniform orbits have to be discretized, for example, by using regular simplices.
	
	In our first paper --- \cite{Radloff:2019:jspi} --- we started with models where the intensity function belongs to the class of monotonous functions. Such models have already been investigated in one dimension, for example, by \cite{Konstantinou:2014} and on multidimensional cuboids or orthants by \cite{Schmidt:2017}. These authors gave the following four conditions on the intensity function~$\lambda$:
\begin{enumerate}[leftmargin=1.5cm]
	\item[$\mathrm{(A1)}$] $\lambda$ is positive on $\mathbb{R}$ and twice continuously differentiable.
	\item[$\mathrm{(A2)}$] The first derivative $\lambda^\prime$ is positive on $\mathbb{R}$.
	\item[$\mathrm{(A3)}$] The second derivative $u^{\prime\prime}$ of $u=\frac{1}{\lambda}$ is injective on $\mathbb{R}$.
	\item[$\mathrm{(A4)}$] The function $\frac{\lambda^\prime}{\lambda}$ is non-increasing.
\end{enumerate}
	Condition $\mathrm{(A2)}$ is the motivation for the name \emph{class of monotonous intensity functions}. The intensity functions of this class have to satisfy always $\mathrm{(A1)}$ to $\mathrm{(A3)}$. $\mathrm{(A4)}$ is an extra condition to guarantee uniqueness.
For a concise notation \[q(x_1)=\lambda(\beta_0+\beta_1 x_1)\] is used and the properties $\mathrm{(A1)}$, $\mathrm{(A2)}$, $\mathrm{(A3)}$ and $\mathrm{(A4)}$ transfer to $q$ for $\beta_1>0$, respectively, and vice versa.
	Poisson regression with intensity function $q_\mathrm{P}(x_1)=\exp(\beta_0+\beta_1 x_1)$ and negative binomial regression 
	as well as special proportional hazard models with censoring, see \cite{Schmidt:2017}, satisfy all four conditions.
			
	If $\beta_1=0$ then the intensity function~$q$ is always a constant. This yields to a (locally) $D$-optimal design as it can be found in linear models. In \citet[section 15.12]{Pukelsheim:1993} such a design consists of the equally weighted vertices of a regular simplex inscribed in the unit sphere, the boundary of the design space. The orientation of the simplex is arbitrary.
			
	The main result for~$\beta_1>0$ in \cite{Radloff:2019:jspi} is recited for the readers' convenience.	
\begin{theorem} \label{Theorem1}
	There is a (locally) $D$-optimal design for the multiple regression problem~\eqref{eq:simplifiedModel} with $\beta_1>0$ and intensity function satisfying $\mathrm{(A1)}$-$\mathrm{(A3)}$ which has one support point equal to $(1,0,\ldots,0)^\top$ and the other $k$~support points are the vertices of an arbitrarily rotated $(k-1)$-dimensional regular simplex which is maximally inscribed in the intersection of the $k$-dimensional unit ball and a hyperplane with $x_1=x_{12}^\ast$.  
	\begin{itemize}
		\item For $k\geq 2$ the position~$x_{12}^\ast\in(-1,1)$ is solution of 
					\begin{equation*}
						\frac{q^\prime(x_{12}^\ast)}{q(x_{12}^\ast)}=\frac{2\,(1+kx_{12}^\ast)}{k\,(1-x_{12}^{\ast\ 2})} \ .
					\label{eq:L9}
					\end{equation*}
					If additionally $\mathrm{(A4)}$ is satisfied, the solution $x_{12}^\ast$ is unique.
		\item For $k=1$ the position~$x_{12}^\ast\in[-1,1)$ is either solution of
					\begin{equation*}
						\frac{q^\prime(x_{12}^\ast)}{q(x_{12}^\ast)}=\frac{2}{1-x_{12}^{\ast}}\ ,
					\label{eq:L92}
					\end{equation*}
					if such a solution exists in~$[-1,1)$, or otherwise~$x_{12}^\ast=-1$.\\
					If additionally $\mathrm{(A4)}$ is satisfied, the solution $x_{12}^\ast$ is unique.
	\end{itemize}
	The design is equally weighted with $\frac{1}{k+1}$.
\end{theorem}
	It should be noted, that for fixed~$\boldsymbol{\beta}$ this theorem does not need~$\mathrm{(A1)}$ to~$\mathrm{(A4)}$ on the entire real line $\mathbb{R}$. It is enough to have it in the ball and so on~$x_1\in[-1,1]$ for~$q$ and on $[\beta_0-\beta_1,\beta_0+\beta_1]$ for~$\lambda$, respectively. But the model has to satisfy the conditions always on the whole real line.

	In our second paper --- \cite{Radloff:2019:moda} --- the conditions~$\mathrm{(A2)}$ and~$\mathrm{(A3)}$ were replaced by~$\mathrm{(A2^\prime)}$ and~$\mathrm{(A3^\prime)}$ and a fifth property~$\mathrm{(A5)}$ was added.
	\begin{enumerate}[leftmargin=1.5cm]
		\item[$\mathrm{(A2^\prime)}$] $\lambda$ is unimodal with mode $c_\lambda^\mathrm{(A2^\prime)}\in\mathbb{R}$.
		\item[$\mathrm{(A3^\prime)}$] There exists a threshold~$c_\lambda^\mathrm{(A3^\prime)}\in\mathbb{R}$ so that the second derivative $u^{\prime\prime}$ of $u=\frac{1}{\lambda}$ is both injective on $(-\infty,c_\lambda^\mathrm{(A3^\prime)}]$ and injective on $[c_\lambda^\mathrm{(A3^\prime)},\infty)$.
		\item[$\mathrm{(A5)}$] $u=\frac{1}{\lambda}$ dominates $z^2$ asymptotically for $z\to\infty$.
	\end{enumerate}		
	In this context condition~$\mathrm{(A2^\prime)}$ means that there exists a $c_\lambda^\mathrm{(A2^\prime)}\in\mathbb{R}$ so that $\lambda^\prime$ is positive on $(-\infty,c_\lambda^\mathrm{(A2^\prime)})$ and negative on $(c_\lambda^\mathrm{(A2^\prime)},\infty)$. Hence, there is only one local maximum which is simultaneously the global maximum. So the class of intensity functions, which satisfy~$\mathrm{(A1)}$, $\mathrm{(A2^\prime)}$ and~$\mathrm{(A3^\prime)}$, is called \emph{class of unimodal intensity functions}.\\	
	Indeed~$\mathrm{(A2)}$ or $\mathrm{(A3)}$ do not imply $\mathrm{(A2^\prime)}$ or $\mathrm{(A3^\prime)}$, respectively. As mentioned before, we only focus on the unit ball and the interval $x_1\in[-1,1]$ for~$q$ or $[\beta_0-\beta_1,\beta_0+\beta_1]$ for~$\lambda$. So in our special case~$\mathrm{(A2)}$ and~$\mathrm{(A3)}$ can be transferred to~$\mathrm{(A2^\prime)}$ and~$\mathrm{(A3^\prime)}$ by using an arbitrary $c_\lambda>\beta_0+\beta_1$, which means that~$c_q$ lies outside the interval~$[-1,1]$ and only one branch of the function is considered.\\	
	Property~$\mathrm{(A5)}$ means \[\lim\limits_{z\to\infty}\left|\frac{u(z)}{z^2}\right|=\infty\ .\] 
	This means that $u(z)=\frac{1}{\lambda(z)}$ goes faster to ($\pm$) infinity than $z^2$ for $z\to\infty$.\\
	As~$\mathrm{(A1)}$ to~$\mathrm{(A4)}$ the conditions~$\mathrm{(A2^\prime)}$, $\mathrm{(A3^\prime)}$ and~$\mathrm{(A5)}$ transfer from the intensity function $\lambda$ to the abbreviated form $q$ for $\beta_1>0$ and vice versa --- analogously $c_q^\mathrm{(\cdot)}=\frac{c_\lambda^\mathrm{(\cdot)}-\beta_0}{\beta_1}$ with $\mathrm{(\cdot)}$ is (A2$^\prime$), (A3$^\prime$) or empty.
	
	The logit model has the intensity function
	\[ q_\mathrm{logit}(x_1)=\frac{\exp(\beta_0+\beta_1 x_1)}{(1+\exp(\beta_0+\beta_1 x_1))^2} \]
	and probit model has
	\[ q_\mathrm{probit}(x_1)=\frac{\phi^2(\beta_0+\beta_1 x_1)}{\Phi(\beta_0+\beta_1 x_1)\cdot(1-\Phi(\beta_0+\beta_1 x_1))} \]
	with the density function $\phi$ and cumulative distribution function $\Phi$ of the standard normal distribution. Both models	satisfy all five conditions~$\mathrm{(A1)}$, $\mathrm{(A2^\prime)}$, $\mathrm{(A3^\prime)}$, $\mathrm{(A4)}$, $\mathrm{(A5)}$ and share a common $c_\lambda^\mathrm{(A2^\prime)}=c_\lambda^\mathrm{(A3^\prime)}=0$, say $c_\lambda=0$. Analogously $c_q=-\frac{\beta_0}{\beta_1}$ for~$q$.\\
	Beside these two models other models like the complementary log-log model, see \cite{Ford:1992}, with intensity function $\lambda_{\mathrm{comp\,log\,log}}(z)=\frac{\exp(2z)}{\exp(\exp(z))-1}$ satisfy all five conditions with $c_\lambda^\mathrm{(A2^\prime)}\approx0.466011$ and $c_\lambda^\mathrm{(A3^\prime)}\approx0.049084$, but here mode~$c_\lambda^\mathrm{(A2^\prime)}$ and threshold~$c_\lambda^\mathrm{(A3^\prime)}$ do not coincide.\\
	We showed that if the (concise) intensity function~$q$ satisfies~$\mathrm{(A1)}$, $\mathrm{(A2^\prime)}$, $\mathrm{(A3^\prime)}$ and $\mathrm{(A5)}$ the (locally) $D$-optimal design~$\xi^\ast=\xi_1^\ast\otimes\overline{\eta}$ is concentrated on exactly two orbits, which are the support points of the marginal design~$\xi_1^\ast$. The idea of the proof is based on~\cite{Biedermann:2006} and~\cite{Konstantinou:2014}.
	
	The next theorem is the main result of our second paper --- \cite{Radloff:2019:moda} --- and is reproduced for the readers' convenience. It characterizes the positions of the two support points of the optimal mar\-gi\-nal design~$\xi_1^\ast$.
\begin{theorem}\label{Theorem2}
	For $k\geq 2$ the simplified problem~\eqref{eq:simplifiedModel} with $\beta_1>0$ and intensity function~$q$ satisfying $\mathrm{(A1)}$, $\mathrm{(A2^\prime)}$, $\mathrm{(A3^\prime)}$ and $\mathrm{(A5)}$ has a (locally) $D$-optimal marginal design~$\xi_1^\ast$ with exactly 2~support points $x_{11}^\ast$ and $x_{12}^\ast$ with $x_{11}^\ast>x_{12}^\ast$ and weights $w_1=\xi_1^\ast(x_{11}^\ast)$ and~$w_2=\xi_1^\ast(x_{12}^\ast)$.\\
	There are 3~cases:
	\begin{enumerate}
		\item[\textit{(a)}] If $c_q^\mathrm{(A2^\prime)}>1$ and $c_q^\mathrm{(A3^\prime)}\notin[-1,1]$, then $x_{11}^\ast=1$, $w_1=\frac{1}{k+1}$, $w_2=\frac{k}{k+1}$ and $x_{12}^\ast\in(-1,1)$ is solution of 
												\begin{equation}
													\frac{q^\prime(x_{12}^\ast)}{q(x_{12}^\ast)}=\frac{2\,(1+kx_{12}^\ast)}{k\,(1-x_{12}^{\ast\ 2})}\ .
													\label{eq:satz:unimodal:Teil:a}
												\end{equation}																			
												If additionally (A4) is satisfied, the solution $x_{12}^\ast$ is unique.\bigskip
		\item[\textit{(b)}] If $c_q^\mathrm{(A2^\prime)}<-1$ and $c_q^\mathrm{(A3^\prime)}\notin[-1,1]$, then $x_{12}^\ast=-1$, $w_1=\frac{k}{k+1}$, $w_2=\frac{1}{k+1}$ and $x_{11}^\ast\in(-1,1)$ is solution of 
												\begin{equation}
													\frac{q^\prime(x_{11}^\ast)}{q(x_{11}^\ast)}=\frac{2\,(-1+kx_{11}^\ast)}{k\,(1-x_{11}^{\ast\ 2})}\ .
													\label{eq:satz:unimodal:Teil:b}
												\end{equation}
												If additionally (A4) is satisfied, the solution $x_{11}^\ast$ is unique.\bigskip											
		\item[\textit{(c)}] Otherwise~$c_q^\mathrm{(A2^\prime)}\in[-1,1]$ or~$c_q^\mathrm{(A3^\prime)}\in[-1,1]$.\\
				Let~$x,y\in\mathbb{R}$ with~$x>y$ and~$\alpha\in\left(-\frac{1}{2},\frac{1}{2}\right)$ be solution of the equation system:
				\begin{align} 		
					\frac{q^\prime(x)}{q(x)}+\frac{2}{x\!-\!y}+(k\!-\!1)\,\frac{q^\prime(x)\,(1\!-\!x^2)\,(\frac{1}{2}\!-\!\alpha) + q(x)\,(-2\,x)\,(\frac{1}{2}\!-\!\alpha)}{q(x)\,(1\!-\!x^2)\,(\frac{1}{2}\!-\!\alpha)+q(y)\,(1\!-\!y^2)\,(\frac{1}{2}\!+\!\alpha)}&=0\label{eqs:allg:x}\\
					\frac{q^\prime(y)}{q(y)}-\frac{2}{x\!-\!y}+(k\!-\!1)\,\frac{q^\prime(y)\,(1\!-\!y^2)\,(\frac{1}{2}\!+\!\alpha) + q(y)\,(-2\,y)\,(\frac{1}{2}\!+\!\alpha)}{q(x)\,(1\!-\!x^2)\,(\frac{1}{2}\!-\!\alpha)+q(y)\,(1\!-\!y^2)\,(\frac{1}{2}\!+\!\alpha)}&=0\label{eqs:allg:y}\\
					\frac{1}{\frac{1}{2}\!-\!\alpha}-\frac{1}{\frac{1}{2}\!+\!\alpha}+(k\!-\!1)\,\frac{q(x)\,(1\!-\!x^2) - q(y)\,(1\!-\!y^2)}{q(x)\,(1\!-\!x^2)\,(\frac{1}{2}\!-\!\alpha)+q(y)\,(1\!-\!y^2)\,(\frac{1}{2}\!+\!\alpha)}&=0 \label{eqs:allg:a}
				\end{align}
				\begin{enumerate}
					\item[\textit{(c0)}] If~$x,y\in(-1,1)$ with~$x>y$ and~$\alpha\in(-\frac{1}{2},\frac{1}{2})$ is a solution of the equation system, the orbit positions are~$x_{11}^\ast=x$, $x_{12}^\ast=y$ with weights~$w_1=\frac{1}{2}-\alpha$ and~$w_2=\frac{1}{2}+\alpha$.
					\item[\textit{(c1)}] If~$x\geq 1$ and~$y\in(-1,1)$, then~$x_{11}^\ast=1$, $w_1=\frac{1}{k+1}$, $w_2=\frac{k}{k+1}$ \linebreak and~$x_{12}^\ast\in(-1,1)$ is the solution of the equation~\eqref{eq:satz:unimodal:Teil:a}.
					\item[\textit{(c2)}] If~$y\leq -1$ and~$x\in(-1,1)$, then~$x_{12}^\ast=-1$, $w_1=\frac{k}{k+1}$, $w_2=\frac{1}{k+1}$ \linebreak and~$x_{11}^\ast\in(-1,1)$ is the solution of the equation~\eqref{eq:satz:unimodal:Teil:b}.
				\end{enumerate}		
	\end{enumerate}
\end{theorem}

\begin{remark}\label{rem:Theorem2:k1}
	Instead of reproducing the whole theorem for $k=1$, only the two main changes in case~\textit{(c)} should be mentioned. So the weights are always~$w_1=w_2=\frac{1}{2}$ and the equation system~\eqref{eqs:allg:x}--\eqref{eqs:allg:a} is replaced by
	\begin{equation}
		\frac{q^\prime(x)}{q(x)}+\frac{2}{x-y}=0 \quad\text{and}\quad \frac{q^\prime(y)}{q(y)}-\frac{2}{x-y}=0\ .
		\label{eqs:allg:k1}
	\end{equation}
\end{remark}

\begin{figure}[t]
	\centering
	\includegraphics[width=6cm]{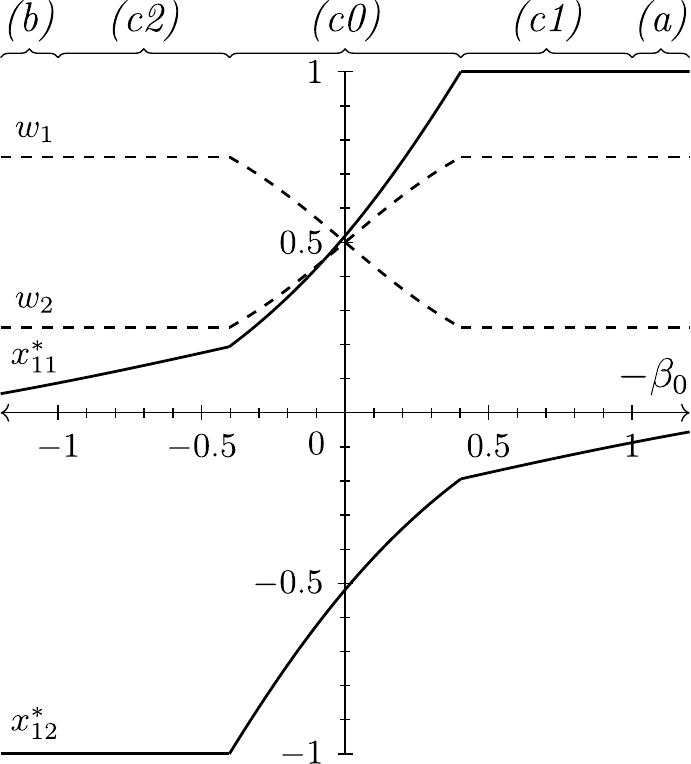}
	\caption{Logit model for $k=3$ and $\beta_1=1$: Dependence of $x_{11}^\ast$ and $x_{12}^\ast$ (solid lines) and the corresponding weights $w_1$ and $w_2=1-w_1$ (dashed lines) on $-\beta_0=-\frac{\beta_0}{\beta_1}=c_q\in[-1.2,1.2]$.}
	\label{fig_logit_k_3_4graphs_2a}
\end{figure}

	To illustrate this complex issue we revisit the logit model in dimension~$k=3$ with $\beta_1=1$. We (numerically) plot the orbit positions~$x_{11}^\ast$ and $x_{12}^\ast$ and corresponding weights~$w_1$ and $w_2=1-w_1$ depending on $-\beta_0=-\frac{\beta_0}{\beta_1}=c_q$, see Figure~\ref{fig_logit_k_3_4graphs_2a}. The cases \textit{(a)} and \textit{(b)} go along with Theorem~\ref{Theorem1} and the results from \cite{Radloff:2019:jspi}. The cases \textit{(c1)} and \textit{(c2)} yield marginal extremum solutions which are identical to \textit{(a)} and \textit{(b)}. So for these four cases there is always an exact minimally supported (locally) $D$-optimal design. As described in Theorem~\ref{Theorem1}, it consists of a pole point in $x_1=-1$ or else $x_1=1$ and the $k$ vertices of a (regular) simplex which is maximally inscribed in the non-degenerated orbit.
	
	But the problematic case is \textit{(c0)} because the (locally) $D$-optimal (generalized) design consists of two non-degenerated orbits and additionally the weights are rarely appropriate for a discretization. In \cite{Radloff:2019:moda} we showed two examples for the logit model ($k=3$, $\beta_1=1$) from which we derived (nearly) exact designs.\\
	For $-\beta_0=0$ the two orbit positions are symmetrical around 0, that is~$x_{11}^\ast=-x_{12}^\ast\approx0.52$, and the weights are~$\xi_1^\ast(x_{11}^\ast)=\xi_1^\ast(x_{12}^\ast)=\frac{1}{2}$. These two orbits were discretized by two 2-dimensional simplices --- overall 6 equally weighted support points; see Figure~\ref{fig_two_balls_1a} (left image).\\
	For $-\beta_0=-0.1$ it is $x_{11}^\ast\approx0.42$, $x_{12}^\ast\approx-0.62$ and $\xi_1^\ast(x_{11}^\ast)\approx0.4297$, while~$0.4297\approx\frac{3}{7}$. We took the rounded design $\xi^\approx$ with the same support points $x_{11}^\ast$ and $x_{12}^\ast$ but with the marginal design $\xi_1^\approx(x_{11}^\ast)=\frac{3}{7}$ and $\xi_1^\approx(x_{12}^\ast)=\frac{4}{7}$. So it was possible to substitute one orbit by the vertices of a 2-dimensional simplex (3~points --- an equilateral triangle) and one by the vertices of a 2-dimensional cube or cross polytope (4~points --- a square). Because of rounding the design $\xi^\approx$ is not optimal but exact and has a high $D$-efficiency, which compares the rounded design $\xi^\approx$ and the optimal design $\xi_{\boldsymbol{\beta}^0}^\ast$ with respect to~$\boldsymbol{\beta}^0$  --- here $p=k+1=4$ and~$\boldsymbol{\beta}^0=(0.1,1,0,0)^\top$:
	\[ \mathrm{Eff}_D(\xi^\approx,\boldsymbol{\beta}^0) = \left(\frac{\det(\boldsymbol{M}(\xi^\approx,\boldsymbol{\beta}^0))}{\det(\boldsymbol{M}(\xi_{\boldsymbol{\beta}^0}^\ast,\boldsymbol{\beta}^0))}\right)^{\!\frac{1}{p}}  \approx 0.999757\ . \]
	
	These designs are not very satisfactory. On the one hand the number of support points is not minimal. On the other hand only special cases have appropriate rational weights which allow a discretization or otherwise the optimality is lost by rounding. Therefore we want to establish minimal supported exact designs for the case \textit{(c0)} in this paper. Mostly these designs wont be optimal but (highly) efficient.
	
	But we start with the reduction of the system of three equations in Theorem~\ref{Theorem2} to only one single equation for special unimodal intensity functions --- symmetrical unimodal intensity functions --- which can be found, for example, in binary response models with logit and probit link.

\section{Optimal Design for Symmetrical Unimodal Intensity Functions}	
\label{sec:symmetric}	
	An interesting observation was made in the discussion section in \cite{Radloff:2019:moda}. For models with unimodal intensity function in which the mode and threshold coincide ($c_\lambda^\mathrm{(A2^\prime)}=c_\lambda^\mathrm{(A3^\prime)}=c_\lambda$) and which are symmetrical, also the two orbit positions are symmetrical in a certain way, which we want to investigate here. For one dimension this has been considered and shown in \citet[Section 6.5 and 6.6]{Ford:1992}, but this proof cannot be extended to higher dimensions directly.
	
\begin{definition}
	An unimodal intensity function in which the mode and threshold coincide ($c_\lambda^\mathrm{(A2^\prime)}=c_\lambda^\mathrm{(A3^\prime)}=c_\lambda$) will be called \emph{symmetrical} to~$c_\lambda$ if 
	\[\lambda(c_\lambda+z)=\lambda(c_\lambda-z)\] 
	for all~$z\in\mathbb{R}$. 
\end{definition}	
The intensity functions of the logit and probit models are symmetrical with~$c_\lambda=0$. But the unimodal intensity function of the complementary log-log model has $c_\lambda^\mathrm{(A2^\prime)}\neq c_\lambda^\mathrm{(A3^\prime)}$ and cannot be symmetrical for this reason.

\begin{lemma}\label{Theorem3}
	Let the intensity function~$\lambda$ be symmetrical to~$c_\lambda$ in the situation of Theorem~\ref{Theorem2}~\textit{(c0)}.
	\begin{itemize}
		\item	For given~$\beta_0\neq c_\lambda$ let $r$ solve 
					\begin{equation} \label{eqs:r:lambdaprimelambda}
						\frac{\lambda^\prime(c_\lambda\!+\!r)}{\lambda(c_\lambda\!+\!r)}=-\,\frac{\splitfrac{-2\,k\,r^2 \left(\beta_1^2\!+\!c^2\!-\!r^2\right)\!+\!\left(\beta_1^2\!-\!c^2\!-\!r^2\right)^2\!-\!4\,c^2\,r^2}{\!+\!\left(\beta_1^2\!-\!c^2\!+\!r^2\right)\sqrt{\left(\beta_1^2\!-\!c^2\!-\!r^2\right)^2\!+\!4\,(k^2\!-\!1)\,c^2\,r^2}}}{(k\!+\!1)\,r\,(r\!+\!c\!-\!\beta_1) (r\!+\!c\!+\!\beta_1) (r\!-\!c\!+\!\beta_1) (r\!-\!c\!-\!\beta_1)}
					\end{equation}		
					with $c:=c_\lambda-\beta_0$. Then 
					\begin{align}
						x&=\frac{c}{\beta_1}+\frac{r}{\beta_1}\ , \label{eqs:r:x}\\
						y&=\frac{c}{\beta_1}-\frac{r}{\beta_1}\ , \label{eqs:r:y}\\
						\alpha&=\frac{-\!\left(\beta_1^2\!-\!c^2\!-\!r^2\right)\!+\!\sqrt{\left(\beta_1^2\!-\!c^2\!-\!r^2\right)^2\!+\!4\,(k^2\!-\!1)\,c^2\,r^2}}{4\,(k\!+\!1)\,c\,r} \label{eqs:r:a}
					\end{align}
					is a solution of the equation system~\eqref{eqs:allg:x}--\eqref{eqs:allg:a}.		
		\item For given~$\beta_0= c_\lambda$ it is~$x=\frac{r}{\beta_1}$, $y=-\frac{r}{\beta_1}$ and $\alpha=0$. Here~$r$ is the solution of
					\begin{equation} \label{eqs:r:lambdaprimelambda2}
						\frac{\lambda^\prime(c_\lambda+r)}{\lambda(c_\lambda+r)}=-\,\frac{2\left(\beta_1^2-k\,r^2\right)}{(k+1)\,r\left(\beta_1^2-r^2\right)}\ .
					\end{equation}	
	\end{itemize}
\end{lemma}

\begin{remark}\label{rem:Theorem3:k1}
	For~$k=1$, see Remark~\ref{rem:Theorem2:k1}, let~$\lambda$ be symmetrical to~$c_\lambda$. Then $x=\frac{c_\lambda-\beta_0}{\beta_1}+\frac{r}{\beta_1}$ and $y=\frac{c_\lambda-\beta_0}{\beta_1}-\frac{r}{\beta_1}$ with~$r$ is solution of 
	\begin{equation}
		\frac{\lambda^\prime(c_\lambda+r)}{\lambda(c_\lambda+r)}=-\frac{1}{r}
	\label{eqs:r:lambdaprimelambda_k1}
	\end{equation}
	solve the equation system~\eqref{eqs:allg:k1}.
\end{remark}


	Lemma~\ref{Theorem3}, whose proof sketch can be found in Appendix~\ref{appendix:proofs}, and Remark~\ref{rem:Theorem3:k1} in combination with Theorem~\ref{Theorem2} give (locally) $D$-optimal designs for models with symmetrical unimodal intensity functions. 
	As a result we reduced the system of equations~\eqref{eqs:allg:x}--\eqref{eqs:allg:a} to only one single equation~\eqref{eqs:r:lambdaprimelambda}. 
	
	But now there is the question if condition $\mathrm{(A4)}$ can guarantee a unique solution as in Theorem~\ref{Theorem1} or in Theorem~\ref{Theorem2}~\textit{(a)} and \textit{(b)} because Theorem~\ref{Theorem2}~\textit{(c)}, especially \textit{(c0)}, tells nothing about uniqueness. But we want to add a remark about the values of~$r$ before.

\begin{remark}\label{brem_sym1}
	Since the system of equations~\eqref{eqs:allg:x}--\eqref{eqs:allg:a} in Theorem~\ref{Theorem2}~\textit{(c0)} should have a solution with two inner support points for the marginal design, $x,y\in(-1,1)$ is required. So
	\[ -1 < \frac{c_\lambda-\beta_0}{\beta_1}\pm\frac{r}{\beta_1} < 1 \]
must be valid.
	This leads with~$\beta_1>0$ to~$r\in\left(-(c_\lambda-\beta_0)-\beta_1,-(c_\lambda-\beta_0)+\beta_1\right)$ and~$r\in\left((c_\lambda-\beta_0)-\beta_1,(c_\lambda-\beta_0)+\beta_1\right)$. Consequently, both intervals must overlap. This happens for~$c_\lambda-\beta_0>0$ at~$0<c_\lambda-\beta_0<\beta_1$ and for~$c_\lambda-\beta_0<0$ at~$-\beta_1<c_\lambda-\beta_0<0$. 
	Thus~$c_\lambda-\beta_0\in(-\beta_1,\beta_1)$ and in particular~$\beta_1^2>(c_\lambda-\beta_0)^2$ must hold. Then~$r$ is in the interval~$\left(|c_\lambda-\beta_0|-\beta_1,-|c_\lambda-\beta_0|+\beta_1\right)$. But Theorem~\ref{Theorem2}~\textit{(c)} need $x>y$ and consequently $r>0$. Hence, $r\in\left(0,-|c_\lambda-\beta_0|+\beta_1\right)$.\\
	This remains valid in particular for~$\beta_0= c_\lambda$, i.~e. $c_\lambda-\beta_0=0$. So~$r\in\left(-\beta_1,\beta_1\right)$. With $r>0$ it is $r\in\left(0,\beta_1\right)$.
\end{remark}

\begin{lemma} \label{Theorem4}
	In situation of Lemma~\ref{Theorem3} let the intensity function~$\lambda$ additionally satisfy condition~$\mathrm{(A4)}$, then equation~\eqref{eqs:r:lambdaprimelambda}, whose right hand side is continuously continued in~$-|c_\lambda-\beta_0|+\beta_1$, has a unique solution in~$r\in\left(0,|c_\lambda-\beta_0|+\beta_1\right)$.\\
	This also holds for~$\beta_0=c_\lambda$ and equation~\eqref{eqs:r:lambdaprimelambda2}, which has exactly one solution in~$r\in\left(0,\beta_1\right)$.
\end{lemma}

\begin{remark}\label{rem:Theorem4:k1}
	For~$k=1$, see Remark~\ref{rem:Theorem3:k1}, and for an intensity function satisfying~$\mathrm{(A4)}$ there is only one solution of~\eqref{eqs:r:lambdaprimelambda_k1}. 
\end{remark}

	The proof sketch of Lemma~\ref{Theorem4} can be found in Appendix~\ref{appendix:proofs}. Lemma~\ref{Theorem4} guarantees a unique solution in~$r\in\left(0,|c_\lambda-\beta_0|+\beta_1\right)$. But Remark~\ref{brem_sym1} points out that for Theorem~\ref{Theorem2}~\textit{(c0)} we need $r\in\left(0,-|c_\lambda-\beta_0|+\beta_1\right)$. This means that the unique solution can result in the two-orbit case or in the one-orbit one-pole case of Theorem~\ref{Theorem2}~\textit{(c)}.


\section{Minimally Supported Designs}
\label{sec:minimal:support}
	In the situation of Theorem~\ref{Theorem1} and Theorem~\ref{Theorem2}~\textit{(a)}, \textit{(b)}, \textit{(c1)} and~\textit{(c2)} the designs have always the minimal number of support points to estimate the parameter vector~$\boldsymbol{\beta}$. These are $k+1$ support points.

\begin{figure}[bh]
	\begin{center}
		\includegraphics[width=0.9\textwidth]{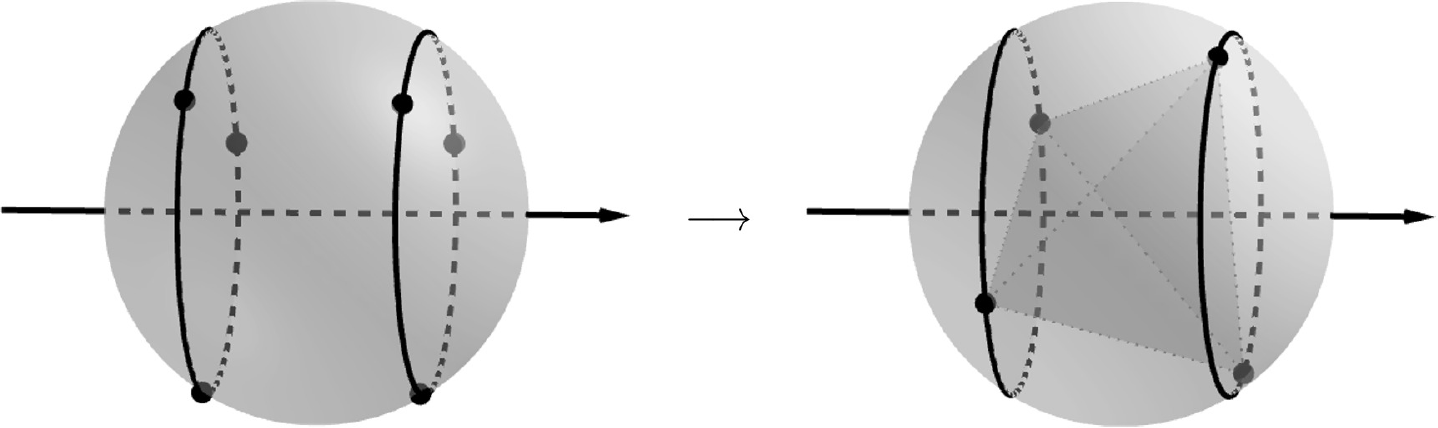} 
	\end{center}
	\caption{Logit model for $k=3$ and $\beta_1=1$ and $-\beta_0=0$: discretized (locally) $D$-optimal designs with 6 or 4 support points.}
	\label{fig_two_balls_1a}
\end{figure}

	In \cite{Radloff:2019:moda} revisited here in the introductory section we indicated exemplarily a (locally) $D$-optimal design for the logit model on the 3-dimensional ball with $-\beta_0=0$ and $\beta_1=1$. This design consists of six support points which are the vertices of two regular 2-dimensional simplices --- equilateral triangles; see Figure~\ref{fig_two_balls_1a} (left image). But this is not the minimum of support points to estimate the four parameters.
	
	So the question arises whether it is possible to reduce the number of support points as it can be found in the concept of \emph{fractional factorial designs}, see, for example, \citet[section 15.11]{Pukelsheim:1993}. Instead of using all vertices of the hypercube~$[-1,1]^k$ as in the full factorial design the fractional factorial design picks only a special percentage of these points. For $k=3$
		\[ (-1,-1,1)^\top,(-1,1,-1)^\top,(1,-1,-1)^\top,(1,1,1)^\top \]
	represent a $2^{3-1}$-fractional factorial design.
	
	In our issue we do not want to pick four of the six points, but we want to use the orthogonality of the spaces spanned by the points (without the $x_1$-component) in the two orbits ($x_1=-1$ and $x_1=1$) of the given $2^{3-1}$-fractional factorial design. Here $\mathrm{span}\{(-1,1)^\top,(1,-1)^\top\}\perp\mathrm{span}\{(-1,-1)^\top,(1,1)^\top\}$. The idea for our problem is illustrated in Figure~\ref{fig_two_balls_1a} (right image). The spanned spaces by points (without the $x_1$-component) in the orbits are orthogonal to each other. And all points span a simplex.
		
	As stated above a (generalized) design~$\xi$ which is rotation invariant with fixed~$x_1$ --- invariant with respect to all orthogonal transformations in~$O(k)$ which do not change the $x_1$-component --- and which has all mass on the unit sphere can be decomposed into a marginal design~$\xi_1$ on $[-1,1]$ and a probability kernel~$\overline{\eta}$ (conditional design), i.~e. $\xi=\xi_1\otimes\overline{\eta}$.	For fixed $x_1$ the kernel $\overline{\eta}(x_1,\cdot)$ is the uniform distribution on the surface of a $(k-1)$-dimensional ball with radius $\sqrt{1-x_1^2}$ --- the orbit at position~$x_1$. If $x_1\in\{-1,1\}$, the $(k-1)$-dimensional ball with the uniform distribution reduces to a single point and represents only a one-point-measure.
	Remembering $q(x_1)=\lambda(\beta_0+\beta_1 x_1)$ the related information matrix, see \cite{Radloff:2019:jspi}, is 
	\begin{equation}
\renewcommand*\arraystretch{1.2}
			\boldsymbol{M}(\xi_1\otimes\overline{\eta},\boldsymbol{\beta}^0)=
					\left(\begin{array}{c|c}
							\hspace*{-0.5em}\begin{array}{cc}
							\int q\,\D\xi_1 		 & \int q \id \D\xi_1\\
							\int q \id \D\xi_1 & \int q \id^2 \D\xi_1
							\end{array} & \mathbb{O}_{2\times (k-1)}\\ \hline
							\mathbb{O}_{(k-1)\times 2} & \frac{1}{k-1} \int q\,(1-\id^2)\,\D\xi_1\ \mathbb{I}_{k-1}							
					\end{array} \right)
	\label{eq:infomatrix}
\renewcommand*\arraystretch{1}
	\end{equation}
	with~$\boldsymbol{\beta}^0=(\beta_0,\beta_1,0,\ldots,0)^\top$.

	The information matrix for a design on the $k$-dimensional unit sphere~$\mathbb{S}_{k-1}$, which is based on exactly two orbits, can be determined analogously to this result. Additionally the uniform distribution does not cover the the full orbits but only sub-spheres.

\begin{lemma} 
\label{L_info_2_orbit_1a}
	Let~$\xi_1$ be the two-point-measure in~$x_{11}$ and~$x_{12}$ with~$\xi_1(x_{11})=\frac{1}{2}-\alpha$ and $\xi_1(x_{12})=\frac{1}{2}+\alpha$ with~$\alpha\in\left(-\frac{1}{2},\frac{1}{2}\right)$. Further let $\overline{\eta}(x_{11},\cdot)$ be a uniform distribution on $\mathbb{S}_{m-2}\bigl(\sqrt{1-\raisebox{0ex}[2ex][0ex]{$x_{11}^2$}}\bigr)\times\left\{0\right\}^{k-m}$ and likewise~$\overline{\eta}(x_{12},\cdot)$ be a uniform distribution on~$\{0\}^{m-1}\times\mathbb{S}_{k-m-1}\bigl(\sqrt{1-\raisebox{0ex}[2ex][0ex]{$x_{12}^2$}}\bigr)$. Then the information matrix is 
		\begin{equation}
	\renewcommand*\arraystretch{1.2}
			\boldsymbol{M}(\xi_1\otimes\overline{\eta},\boldsymbol{\beta}^0) =
						\left(\begin{array}{c|c}
								\hspace*{-0.5em}\begin{array}{cc}
								\int q\,\D\xi_1 		 & \int q \id \D\xi_1\\
								\int q \id \D\xi_1 & \int q \id^2 \D\xi_1
								\end{array} & \mathbb{O}_{2\times (k-1)}\\ \hline
								\mathbb{O}_{(k-1)\times 2} &   \begin{array}{cc}
																										c_1\, \mathbb{I}_{m-1} & \mathbb{O}_{(m-1)\times(k-m)} \\
																										\mathbb{O}_{(k-m)\times(m-1)} & c_2\, \mathbb{I}_{k-m}								
																								\end{array}								
						\end{array} \right) 
		\label{eq:infomatrix_2_orbits}
	\renewcommand*\arraystretch{1}
		\end{equation}
	with $c_1=\frac{1}{m-1}\,q(x_{11}) \,(1\!-\!x_{11}^2) \,(\frac{1}{2}\!-\!\alpha)$ and $c_2=\frac{1}{k-m}\,q(x_{12}) \,(1\!-\!x_{12}^2) \,(\frac{1}{2}\!+\!\alpha)$.
\end{lemma}	

	Now the optimality case in Theorem~\ref{Theorem2}~\textit{(c0)} on two orbits should be used to investigate when both information matrices~\eqref{eq:infomatrix} und~\eqref{eq:infomatrix_2_orbits} are identical. With that both related (generalized) designs would be (locally) $D$-optimal.

\begin{lemma}\label{bT3}
	Both information matrices~\eqref{eq:infomatrix} and~\eqref{eq:infomatrix_2_orbits} are identical in the situation of Theorem~\ref{Theorem2}~\textit{(c0)} if and only if $\alpha=\frac{1}{2}-\frac{m}{k+1}$.
\end{lemma}

	The proof can be found in Appendix~\ref{appendix:proofs}.

	Consequently both orbits need the weights~$\xi_1(x_{11})=\frac{m}{k+1}$ and~$\xi_1(x_{12})=\frac{k-m+1}{k+1}$ to coincide both information matrices. This allows an experimental design, which has the same value for the $D$-optimality criterion, consisting of two orbits with~$m$ and with~$k-m+1$~support points. This can be done by two regular simplices --- one simplex in dimension~$m-1$ and one in dimension~$k-m$. So the simplices are the discretizations of the uniform distributions on~$\mathbb{S}_{m-2}\bigl(\sqrt{1-x_{11}^2}\bigr)\times\left\{0\right\}^{k-m}$ and on~$\{0\}^{m-1}\times\mathbb{S}_{k-m-1}\bigl(\sqrt{1-x_{12}^2}\bigr)$.
	
	Let $\boldsymbol{S}_m\in\mathbb{R}^{m\times (m+1)}$ be a matrix, where the columns represent the $m+1$ vertices of an $m$-dimensional regular simplex (in~$\mathbb{R}^m$). Then the columns of the matrix
	\begin{equation*}
	\renewcommand*\arraystretch{1.2}
					\left(\begin{array}{c|c}
							x_{11} \mathds{1}_{m}^\top & x_{12} \mathds{1}_{k-m+1}^\top\\ \hline
							\boldsymbol{R}_1\,\boldsymbol{S}_{m-1} & \mathbb{O}_{(m-1)\times (k-m+1)} \\ \hline
							\mathbb{O}_{(k-m)\times m}	&	\boldsymbol{R}_2\,\boldsymbol{S}_{k-m} 
					\end{array} \right)
\renewcommand*\arraystretch{1}
	\end{equation*}
with arbitrary orthogonal transformations~$\boldsymbol{R}_1\in O(m-1)$ and~$\boldsymbol{R}_2\in O(k-m)$ represent the support points of such a minimal supported design.
	\[  \left(\left.\sqrt{\frac{m+1}{m}}\,\mathbb{I}_m + \frac{1-\sqrt{m+1}}{m \sqrt{m}}\,\mathds{1}_m \mathds{1}_m^\top \right| -\frac{1}{\sqrt{m}}\,\mathds{1}_m \right) \in\mathbb{R}^{m\times (m+1)} \]
	is an example for $\boldsymbol{S}_m$. In this notation $\mathbb{I}_m$ stands for the standard simplex which needs to be scaled and shifted appropriately so that it is in combination with the last vertex $-\frac{1}{\sqrt{m}}\,\mathds{1}_m$ (last column) a regular simplex on the unit sphere~$\mathbb{S}_{m-1}$.

\begin{figure}[t]
\centering
\includegraphics[height=7cm]{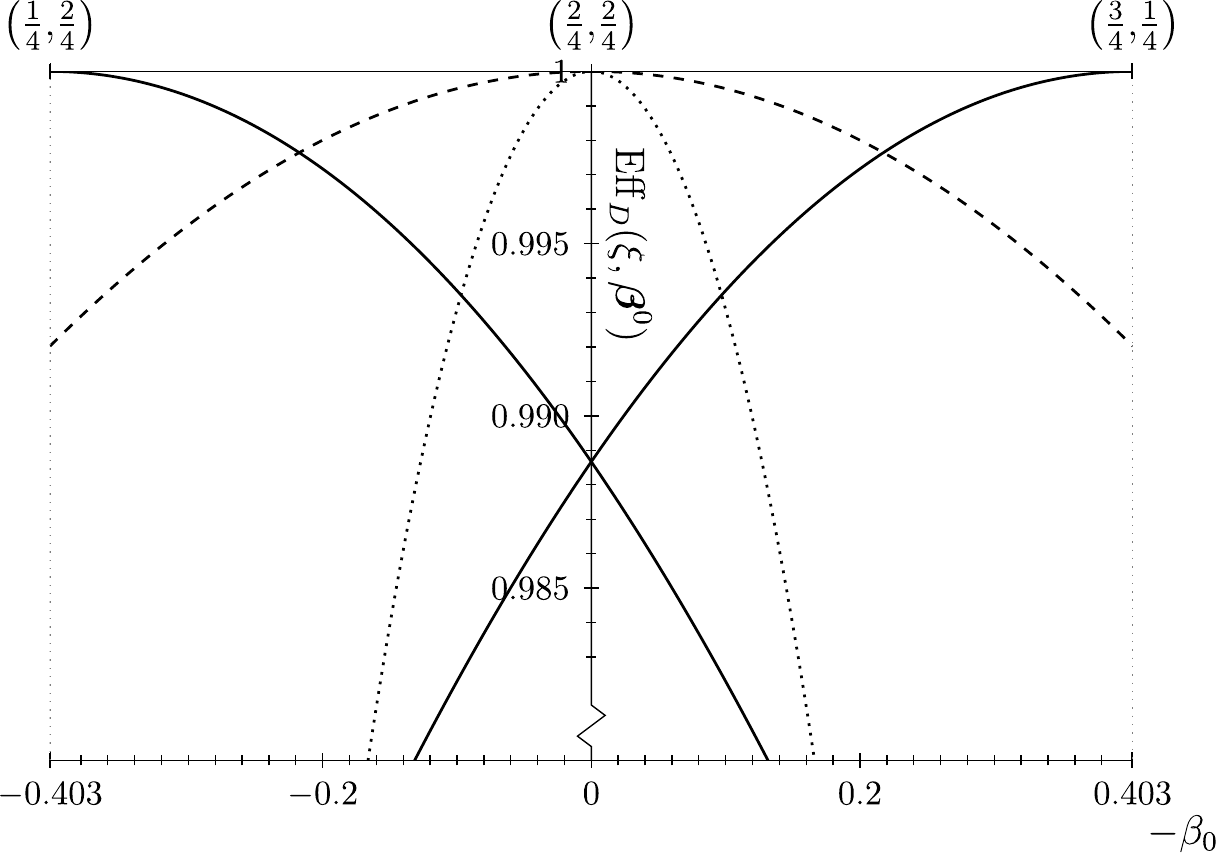}
\caption{$D$-efficiency for the logit model with $k=3$ and $\beta_1=1$: comparison of designs with exactly $k+1=4$ equally weighted support points in $-\beta_0\in(-0.403, 0.403)$ (rounded).}
\label{fig_eff_k3_1a}
\end{figure}

	Finally, we want to look at the $D$-efficiency, here with~$\boldsymbol{\beta}^0=(\beta_0,\beta_1,0,\ldots,0)^\top$,
	\[ \mathrm{Eff}_D(\xi,\boldsymbol{\beta}^0) = \left(\frac{\det(\boldsymbol{M}(\xi,\boldsymbol{\beta}^0))}{\det(\boldsymbol{M}(\xi_{\boldsymbol{\beta}^0}^\ast,\boldsymbol{\beta}^0))}\right)^{\!\frac{1}{p}}  \in[0,1] \]
	for designs~$\xi$ with exactly $p=k+1$ equally weighted support points in the region where two non-degenerated orbits occur.

\begin{figure}[t]
\centering
\includegraphics[height=7cm]{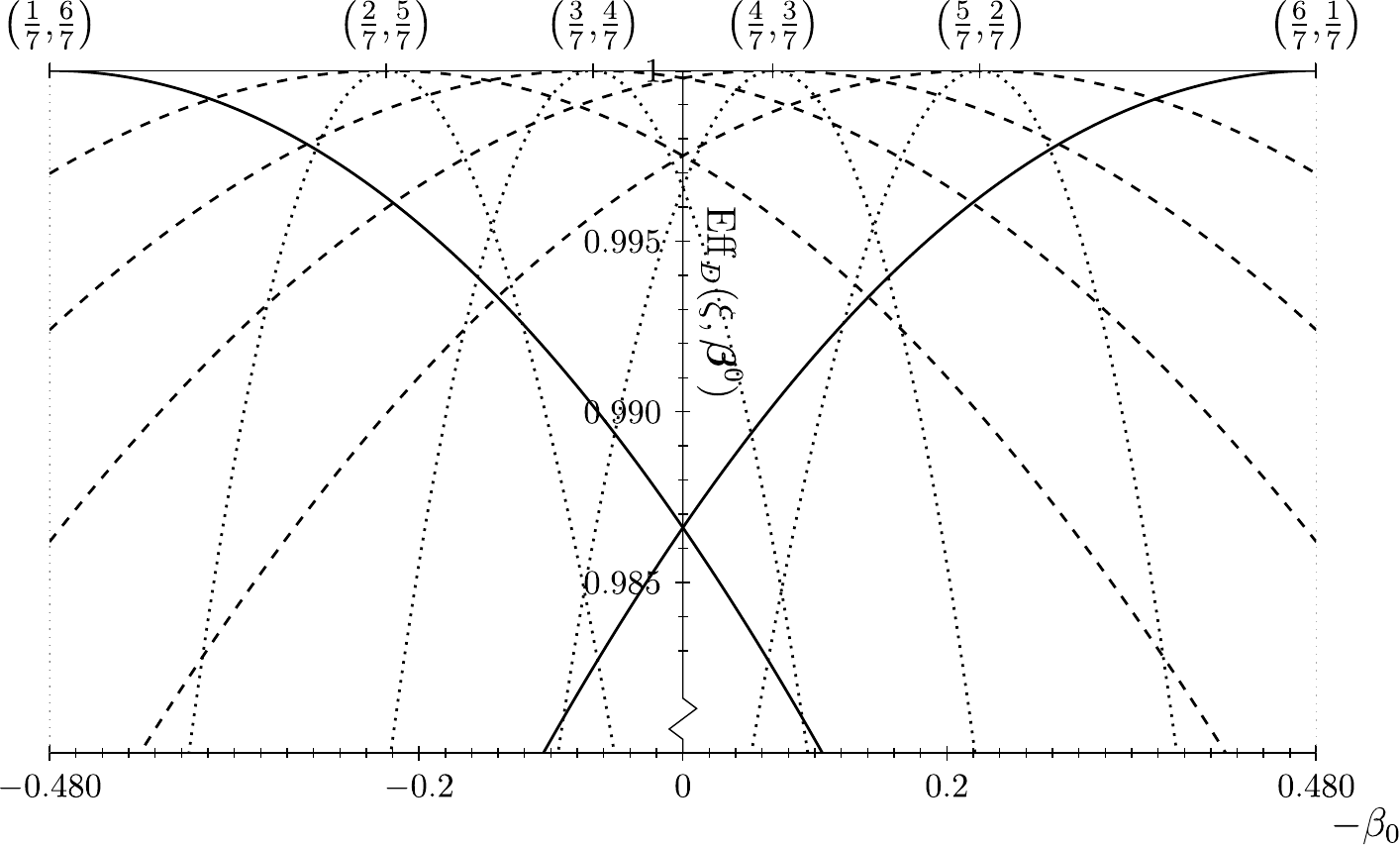}
\caption{$D$-efficiency for the logit model with $k=6$ and $\beta_1=1$: comparison of designs with exactly $k+1=7$ equally weighted support points in $-\beta_0\in(-0.480, 0.480)$ (rounded).}
\label{fig_eff_k6_1a}
\end{figure}

	As an example, the logit model with~$\beta_1=1$ is used to determine the $D$-efficiency in dimensions~$k=3$ and~$k=6$. In Figure~\ref{fig_eff_k3_1a} and Figure~\ref{fig_eff_k6_1a} only the regions for $-\beta_0$ with two non-degenerated orbits in the optimal design (case \textit{(c0)} in Theorem~\ref{Theorem2}), i.~e.~$-\beta_0\in(-0.403,0.403)$ (rounded) for~$k=3$ and~$-\beta_0\in(-0.480,0.480)$ (rounded) for~$k=6$, are plotted.
	
	For this purpose, three different types of exact designs are compared with the (locally) $D$-optimal design~$\xi_{\boldsymbol{\beta}^0}^\ast$. The optimal design is a generalized design with real weights. Therefore it cannot be discretized as an exact design in general.
	
	First, the two optimal exact designs with one pole and one orbit, which are discretized as a regular $(k-1)$-dimensional simplex, are used for comparison. The orbit position remains unchanged and is determined at the boundary values~$-\beta_0\approx\pm0.403$ or~$-\beta_0\approx\pm0.480$. See the solid lines in both figures. 
	
	Second, the designs with the same orbit position as the associated design which is (locally) optimal for~$-\beta_0$ are the next alternative. Only the weights were rounded/shifted to integral multiples of~$\frac{1}{k+1}$. See the dotted lines.
	
	Third, the designs with fixed design weights which are integral multiples of~$\frac{1}{k+1}$ represent the last model category. So only the positions of the orbits have to be optimized with these fixed design weights. This can be done by solving only the equations~\eqref{eqs:allg:x} and~\eqref{eqs:allg:y} with the selected weights in Theorem~\ref{Theorem2}~\textit{(c)}. Equation~\eqref{eqs:allg:a} is omitted. See the dashed lines in both plots.
	
	The Figure~\ref{fig_eff_k3_1a} reveals for dimension~$k=3$ that there are only three positions in the range~$-\beta_0\in[-0.403,0.403]$ (rounded) where (locally) $D$-optimal designs with the minimal number of support points --- four points --- exists.	
	For~$-\beta_0\approx-0.403$ this is the design consisting of the pole~$x_{12}^\ast=-1$ and one orbit at~$x_{11}^\ast$ with three points as vertices of an equilateral triangle. Then for~$-\beta_0=0$ there are two orbits with two points each. And, at~$-\beta_0\approx 0.403$ the design consists of one orbit at~$x_{12}^\ast$ with three equally weighted support points and the pole~$x_{11}^\ast=1$.
	In the span between these optimality positions the considered discretizations provide a fairly high efficiency. Using the transition  directly from pole and orbit to orbit and pole, the efficiency is always greater than~$0.988$ (intersection of the solid lines). If the two orbits are also discretized in between, the efficiency is greater than~$0.993$ (intersection of dotted line and solid lines) or even greater than~$0.997$ (intersection of dashed line and solid lines).

	For dimension~$k=6$, see figure~\ref{fig_eff_k6_1a}, an efficiency of more than~$0.986$ is possible by stepping directly from pole and orbit with six support points to orbit with six design points and pole. If the intermediate steps --- two orbits with~2 and 5~points, 3~and 4~points, 4~and 3~points as well as 5~and 2~points --- are used, then by simple rounding of the weights to integral multiples of~$\frac{1}{k+1}$ an efficiency greater than~$0.995$ (dotted lines) and with additional optimization of the orbit positions even greater than~$0.999$ (dashed lines) can be achieved.

\section{Conclusion}
	In summary it can be postulated that very efficient designs are generated based on only $k+1$~design points which is the minimal number of support points to estimate the parameter vector. It seems that higher dimensions enable designs with higher $D$-efficiency, in particular using the third option of discretization. Here we only considered designs with exactly two orbits. Thus it cannot be excluded that there are designs with a better efficiency or even (locally) optimal designs which are supported by exactly $k+1$~points. Maybe these designs have support points which lie not on the orbit but are jittered a little bit. This as well as a potential lower efficiency bound needs further investigations.
	
	On the other side the reduction of the equation system to one single equation for determining (locally) $D$-optimal design for symmetrical unimodal intensity functions is a nice feature and can help to decrease computing costs. 
	
	Also the question of optimal designs on the ball with respect to other optimality criteria should be considered in future. 
	
	Finally, we want to emphasize that the established designs do not only work for the unit ball. By using the concept of equivariance for linear transformations, say scaling, reflecting and rotating, the class of design spaces can be extended to $k$-dimensional balls with arbitrary radius or any $k$-dimensional ellipsoid.


\begin{appendices}

\section{Notation}
	\begin{center}
		\begin{tabular}{ll}
			$\mathbb{B}_k$ & $k$-dimensional unit ball\\
			$\mathbb{B}_k(r)$ & $k$-dimensional ball with radius $r$\\
			$\mathbb{S}_{k-1}$ & unit sphere, which is the surface of~$\mathbb{B}_k$\\
			$\mathbb{S}_{k-1}(r)$ & sphere with radius~$r$, which is the surface of~$\mathbb{B}_k(r)$\\
			$\mathbb{O}_k$ & $k$-dimensional zero-vector\\
			$\mathbb{O}_{k_1\times k_2}$ & $(k_1\times k_2)$-dimensional zero-matrix\\
			$\mathds{1}_k$ & $k$-dimensional one-vector\\
			$\mathbb{I}_k$ & $(k\times k)$-dimensional identity matrix\\
			$\id$ & identity function		
		\end{tabular}
	\end{center}

\section{Proofs}\label{appendix:proofs}

\begin{proof}[\textbf{Proof sketch of Lemma~\ref{Theorem3}}]
	By plugging~\eqref{eqs:r:x} and~\eqref{eqs:r:y} into~\eqref{eqs:allg:a} and using the symmetry to simplify, we get
	\begin{equation*}
		\frac{-2\,\alpha \left(4\,c\,r\,\alpha\!+\!\left(\beta_1^2\!-\!c^2\!-\!r^2\right)\right)\!+\!4\,(k\!-\!1)\,c\,r\left(\frac{1}{2}\!-\!\alpha\right) \left(\frac{1}{2}\!+\!\alpha\right)}{\left(\frac{1}{2}\!-\!\alpha\right) \left(\frac{1}{2}\!+\!\alpha\right)\left(4\,c\,r\,\alpha\!+\!\left(\beta_1^2\!-\!c^2\!-\!r^2\right)\right)}=0\ .
	\end{equation*}	
	In the numerator there is a polynomial of degree two in~$\alpha$ with the two roots~$\alpha_{\mp}(r)$ depending on~$r$:
	\begin{equation*}
		\alpha_{\mp}(r):=\frac{-\left(\beta_1^2-c^2-r^2\right)\mp\sqrt{\left(\beta_1^2-c^2-r^2\right)^2+4\,(k+1)\,(k-1)\,c^2\,r^2}}{4\,(k+1)\,c\,r}\ .
	\end{equation*}
	Now we examine the values of~$\alpha_{\mp}(r)$ depending on~$r$. Only $-|c|-\beta_1$, $|c|-\beta_1$, $-|c|+\beta_1$ or $|c|+\beta_1$ can solve the expression~$\alpha_{\mp}(r)=\pm\frac{1}{2}$. But~$-|c|-\beta_1$ and~$|c|+\beta_1$ are not in the interesting region for~$r$. We have 
	\begin{equation*}
		\alpha_{-}\left(\pm(|c|-\beta_1)\right)=\pm\frac{1}{2}\,\sign(c) \quad\text{and}\quad 
		\alpha_{+}\left(\pm(|c|-\beta_1)\right)=\mp\frac{1}{2}\,\sign(c)\, \frac{k-1}{k+1}\ .
	\end{equation*}	
	Because of $\lim_{r\nearrow 0}\alpha_{-}\left(r\right)=\sign(c)\infty$ and $\lim_{r\searrow 0}\alpha_{-}\left(r\right)=-\sign(c)\infty$ the root $\alpha_{-}(r)$ has in the interval~$r\in\left[|c|-\beta_1,-|c|+\beta_1\right]$ only values outside~$(-\frac{1}{2},\frac{1}{2})$. Hence, $\alpha_{-}(r)$ is not a relevant root.\\
	Since $\lim_{r\to 0}\alpha_{+}\left(r\right)=0$	the discontinuity of the root~$\alpha_{+}(r)$ in~$r=0$ can be removed. So~$\alpha_{+}(r)$ has only values in~$(-\frac{1}{2},\frac{1}{2})$ on the interval~$r\in\left[|c|-\beta_1,-|c|+\beta_1\right]$ and~$\alpha_{+}(r)$, which is~\eqref{eqs:r:a}, is the only relevant root.\\
	After inserting~\eqref{eqs:r:x} and~\eqref{eqs:r:y} into~\eqref{eqs:allg:x} as well as~\eqref{eqs:r:x} and~\eqref{eqs:r:y} into~\eqref{eqs:allg:y} and subtracting both obtained equations and simplifying by using the symmetry, we get
	\begin{equation*}
		\frac{(k+1)\,\lambda^\prime(c_\lambda+r)}{\lambda(c_\lambda+r)}=-(k-1)\,\frac{-2\,r+\alpha\cdot 4\,c}{ \left(\beta_1^2-c^2-r^2\right) +\alpha\cdot 4\,c\,r}-\frac{2}{r}	\ .
	\end{equation*}		
	Equation~\eqref{eqs:r:lambdaprimelambda} follows by plugging~$\alpha_+(r)$ as~$\alpha$ into it and by some simplifications.
	
	For $\beta_0=c_\lambda$, i.~e. $c=c_\lambda-\beta_0=0$, we get directly~$\alpha=0$ by inserting $x=\frac{r}{\beta_1}$ and $y=-\frac{r}{\beta_1}$ in~\eqref{eqs:allg:a} and exploiting the symmetry. This is inserted in~\eqref{eqs:allg:x} and in~\eqref{eqs:allg:y}. The difference between these two equations results in~\eqref{eqs:r:lambdaprimelambda2}.
\end{proof}

\begin{proof}[\textbf{Proof sketch of Lemma~\ref{Theorem4}}]
	This proof is a lot of curve sketching. We start with~$\beta_0\neq c_\lambda$. The denominator of the right hand side of~\eqref{eqs:r:lambdaprimelambda} has five roots in~$r$. $-|c_\lambda-\beta_0|-\beta_1<0$ and $|c_\lambda-\beta_0|-\beta_1<0$ are not in the considered interval~$\left(0,|c_\lambda-\beta_0|+\beta_1\right)$. In $r=-|c_\lambda-\beta_0|+\beta_1$ there is a discontinuity which can be removed. In~$r=0$ and in $r=|c_\lambda-\beta_0|+\beta_1$ there are two poles. Analyzing these poles for the considered interval we see that the values start from~$-\infty$ ($r\searrow 0$) and go up to~$+\infty$ ($r\nearrow |c_\lambda-\beta_0|+\beta_1$). Sophisticated curve sketching shows that the right hand side of~\eqref{eqs:r:lambdaprimelambda} is strictly monotonically increasing on~$\left(0,|c_\lambda-\beta_0|+\beta_1\right)$. So it is strictly monotonically increasing and covers~$(-\infty,\infty)$. In combination with~$\mathrm{(A4)}$ for the left hand side of~\eqref{eqs:r:lambdaprimelambda} (monotonically decreasing) there is exactly one solution.
	
	For~$\beta_0=c_\lambda$ we can mention that the right hand side of~\eqref{eqs:r:lambdaprimelambda2} is also strictly monotonically increasing on~$(0,\beta_1)$. Hence, there is only one solution.
	
	An analogue result holds for the situation in Remark~\ref{rem:Theorem4:k1}.
\end{proof}

\begin{proof}[\textbf{Proof of Lemma~\ref{bT3}}]
	Rearranging equation~\eqref{eqs:allg:a} equivalently in two ways gives
	\begin{align*}
		q(x_{12})\,(1\!-\!x_{12}^2)\,(\tfrac{1}{2}\!+\!\alpha) &=	q(x_{11})\,(1\!-\!x_{11}^2)\,(\tfrac{1}{2}\!-\!\alpha) \, \frac{k\,(\tfrac{1}{2}\!+\!\alpha)\!-\!(\tfrac{1}{2}\!-\!\alpha) }{ k\,(\tfrac{1}{2}\!-\!\alpha)\!-\!(\tfrac{1}{2}\!+\!\alpha) }  	\text{\quad and}\\ 
		q(x_{11})\,(1\!-\!x_{11}^2)\,(\tfrac{1}{2}\!-\!\alpha) &= q(x_{12})\,(1\!-\!x_{12}^2)\,(\tfrac{1}{2}\!+\!\alpha)\, \frac{ k\,(\tfrac{1}{2}\!-\!\alpha)\!-\!(\tfrac{1}{2}\!+\!\alpha) }{k\,(\tfrac{1}{2}\!+\!\alpha)\!-\!(\tfrac{1}{2}\!-\!\alpha) } \ .
	\end{align*}
	The two denominators are zero if and only if $\alpha=\frac{1}{2}-\frac{1}{k+1}$ and $\alpha=\frac{1}{2}-\frac{k}{k+1}$, respectively. But this cannot happen to non-degenerated orbits because $\frac{1}{2}-\frac{k}{k+1}<\alpha<\frac{1}{2}-\frac{1}{k+1}$.\\
	Putting both equations into the diagonal entry of the information matrix~\eqref{eq:infomatrix} yield
	\begin{align*}
		\frac{1}{k-1} &\int q\,(1-\id^2)\,\D\xi_1 \\
		&= q(x_{11}) \,(1\!-\!x_{11}^2) \,(\tfrac{1}{2}\!-\!\alpha) \left[ \frac{1}{k-1}  + \frac{1}{k-1}\cdot   \frac{k\,(\tfrac{1}{2}\!+\!\alpha)\!-\!(\tfrac{1}{2}\!-\!\alpha) }{ k\,(\tfrac{1}{2}\!-\!\alpha)\!-\!(\tfrac{1}{2}\!+\!\alpha) } \right]
	\intertext{and}
		\frac{1}{k-1} &\int q\,(1-\id^2)\,\D\xi_1 \\
		&= q(x_{12})\,(1\!-\!x_{12}^2)\,(\tfrac{1}{2}\!-\!\alpha) \left[ \frac{1}{k-1}\cdot \frac{ k\,(\tfrac{1}{2}\!-\!\alpha)\!-\!(\tfrac{1}{2}\!+\!\alpha) }{k\,(\tfrac{1}{2}\!+\!\alpha)\!-\!(\tfrac{1}{2}\!-\!\alpha) }  +   \frac{1}{k-1} \right] 
	\end{align*}	
	They are identical to the diagonal entries of the information matrix~\eqref{eq:infomatrix_2_orbits} in Lemma~\ref{L_info_2_orbit_1a} if and only if
	\[ \frac{1}{k\!-\!1}  + \frac{1}{k\!-\!1}\cdot   \frac{k\,(\tfrac{1}{2}\!+\!\alpha)\!-\!(\tfrac{1}{2}\!-\!\alpha) }{ k\,(\tfrac{1}{2}\!-\!\alpha)\!-\!(\tfrac{1}{2}\!+\!\alpha) }  \!=\! \frac{1}{m\!-\!1} \text{\ and\ }
			\frac{1}{k\!-\!1}\cdot \frac{ k\,(\tfrac{1}{2}\!-\!\alpha)\!-\!(\tfrac{1}{2}\!+\!\alpha) }{k\,(\tfrac{1}{2}\!+\!\alpha)\!-\!(\tfrac{1}{2}\!-\!\alpha) }  +   \frac{1}{k\!-\!1} \!=\! \frac{1}{k\!-\!m} 	
	\]
	which are both equivalent to $\alpha = \frac{1}{2}-\frac{m}{k+1}$.
\end{proof}
	
\end{appendices}




%
%
%
%

\end{document}